\documentclass[graybox]{svmult}
\usepackage{multirow}
\usepackage{upgreek}
\usepackage[]{apacite}

\usepackage[square, numbers]{natbib}

\usepackage{type1cm}        
\usepackage{makeidx}         
\usepackage{graphicx}        
\usepackage{multicol}        
\usepackage[bottom]{footmisc}

\usepackage{newtxtext}       %
\usepackage{newtxmath}       
\DeclareFontEncoding{FML}{}{}%
\DeclareFontSubstitution{FML}{fncmi}{m}{it}%
\DeclareSymbolFont{fouriernc}{FML}{fncmi}{m}{it}%
\DeclareMathAccent{\fvec}{0}{fouriernc}{"7E}

\makeindex             

\authorrunning{L. Bugnet et al.}
\titlerunning{The impact of a fossil magnetic field on dipolar mixed-mode frequencies}

\begin{document}

\title*{The impact of a fossil magnetic field on dipolar mixed-mode frequencies in sub- and red-giant stars}
\author{L. Bugnet, V. Prat, S. Mathis, R. A. García, S. Mathur, K. Augustson, C. Neiner, and  M. J. Thompson$^\dagger$}
\institute{L. Bugnet, V. Prat, S. Mathis, R.A. Garc\'\i a, and K. Augustson \at AIM, CEA, CNRS, Université Paris-Saclay, Université Paris Diderot, Sorbonne Paris Cité, F-91191 Gif-sur-Yvette, France \email{lisa.bugnet@cea.fr}
\and S. Mathur \at Instituto de Astrof\'{\i}sica de Canarias, E-38200, La Laguna, Tenerife, Spain
\and C. Neiner \at LESIA, Paris Observatory, CNRS, PSL University, Sorbonne Université, Univ. Paris Diderot, Sorbonne Paris Cité, 5 place Jules Janssen, 92195 Meudon, France
\and M. J. Thompson$^\dagger$}
%
%
\maketitle

\abstract*{Each chapter should be preceded by an abstract (no more than 200 words) that summarizes the content. The abstract will appear \textit{online} at \url{www.SpringerLink.com} and be available with unrestricted access. This allows unregistered users to read the abstract as a teaser for the complete chapter.
Please use the 'starred' version of the \texttt{abstract} command for typesetting the text of the online abstracts (cf. source file of this chapter template \texttt{abstract}) and include them with the source files of your manuscript. Use the plain \texttt{abstract} command if the abstract is also to appear in the printed version of the book.}

\abstract{The recent discovery of low-amplitude dipolar oscillation mixed modes in massive red giants indicate the presence of a missing physical process inside their cores. Stars more massive than $\sim 1.3$ M$_\odot$ are known to develop a convective core during the main-sequence: the dynamo process triggered by this convection could be the origin of a strong magnetic field inside the core of the star, trapped when it becomes stably stratified and for the rest of its evolution. The presence of highly magnetized white dwarfs strengthens the hypothesis of buried fossil magnetic fields inside the core of evolved low-mass stars. If such a fossil field exists, it should affect the mixed modes of red giants as they are sensitive to processes affecting the deepest layers of these stars. The impact of a magnetic field on dipolar oscillations modes was one of Pr. Michael J. Thompson's research topics during the 90s when preparing the helioseismic SoHO space mission. As the detection of gravity modes in the Sun is still controversial, the investigation of the solar oscillation modes did not provide any hint of the existence of a magnetic field in the solar radiative core. Today we have access to the core of evolved stars thanks to the asteroseismic observation of mixed modes from CoRoT, \textit{Kepler}, K2 and TESS missions. The idea of applying and generalizing the work done for the Sun came from discussions with Pr. Michael Thompson in early 2018 before we loss him. Following the path we drew together, we theoretically investigate the effect of a stable axisymmetric mixed poloidal and toroidal magnetic field, aligned with the rotation axis of the star, on the mixed modes frequencies of a typical evolved low-mass star. This enables us to estimate the magnetic perturbations to the eigenfrequencies of mixed dipolar modes, depending on the magnetic field strength and the evolutionary state of the star. We conclude that strong magnetic fields of $\sim$ 1MG should perturbe the mixed-mode frequency pattern enough for its effects to be detectable inside current asteroseismic data.
}

\section{Introduction: understanding red giant dynamics}
\label{sec:1}

Amongst key observations of red-giant (RG) interiors that stands out are their interestingly low core rotation rate \citep[e.g.][]{beck2011, Mosser2012, Gehan2018} that is only about 10 times higher than the surface rate. However, state-of-the-art stellar evolution models predict that the core should rotate about 100 times faster than the stellar surface according to current transport models \citep[e.g.][]{Eggenberger2012b, Ceillier2013, Marques2013a, Cantiello2014}. This highlights an strong bottleneck concerning the understanding of the strength of the angular momentum transport inside evolved Solar-like stars. This is complemented by the surprising discovery of two populations of RG where the character of their oscillations depends upon their masses \citep[e.g.][]{garcia2014b, Stello2016a, Mosser2017a}. On the one hand, the lower mass (1 M$_\odot$ < M$_\star$ < 1.3 M$_\odot$) range of stars exhibit solar-like oscillations with normal amplitude. On the other hand, nearly 50\% of the more massive group (1.5 M$_\odot$ < M$_\star$ < 2.5 M$_\odot$) exhibit some mixed modes with very low amplitudes. This emphasises that some oscillation modes are weakened and sometimes they disappear in the core of a large fraction of RG. Once again, the current simplified theory needs to be modified to account for the origin of these weakened modes.\\

\begin{figure}[t]
\sidecaption
\includegraphics[scale=.35]{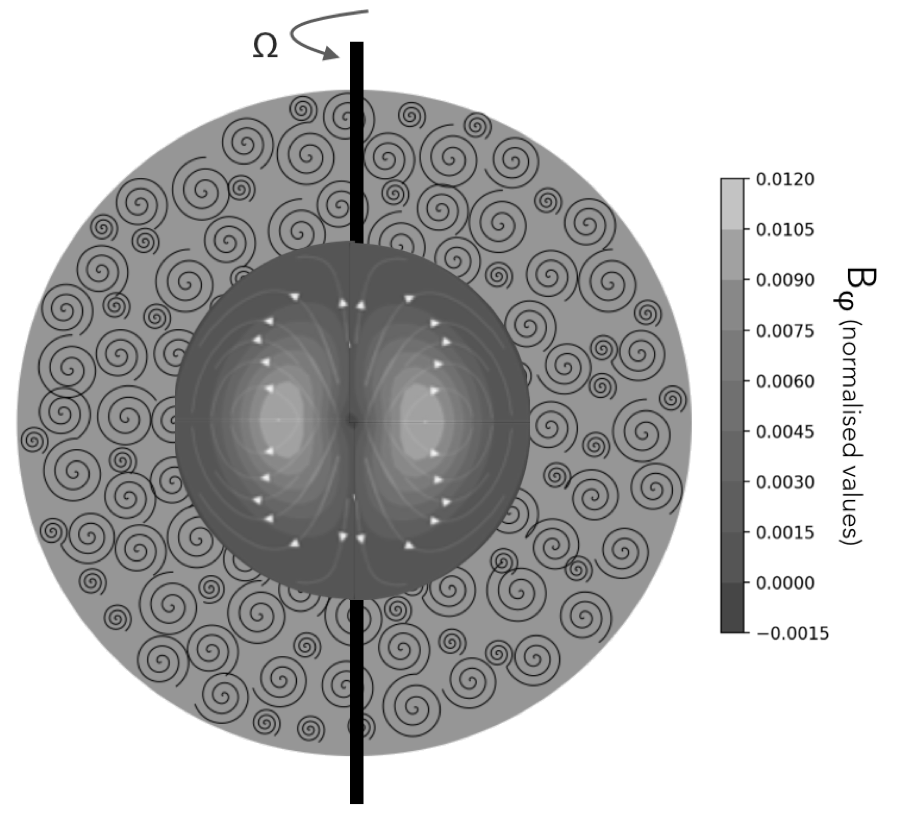}
\caption{Representation of a trapped magnetic field inside an RG core. The magnetic field has a poloidal (white lines) and a toroidal (grey scale) components normalised by the strength $B_0$ of the field computed by using the semi-analytic description of \cite{Duez2010, Duez2010a}.}
\label{fig:1}       
\end{figure}

 The idea of a strong stable buried magnetic field altering the oscillation modes of red giants (RG) came from \cite{Fuller2015a} who invoked a conversion from gravity waves towards magneto-gravity waves that would remain trapped inside the core of the star. This theory is controversial \citep{Mosser2017a}, but has been systematically studied and improved \citep{Cantiello2016, Loi2017a, Lecoanet2017b}, revealing that magnetism is still a very good candidate for both angular momentum transport \citep{Fuller2019, Eggenberger2019} and mode supression. It was shown in \cite{Braithwaite2004} that the relaxation of a stochastic field in a radiative zone leads to the formation of a stable, fossil, mixed poloidal and toroidal magnetic field. The energy minimization principle employed in \cite{Duez2010} yields analytical approximations to such a mixed field solution. We use this configuration as provided in \cite{Duez2010a} to model a fossil magnetic field inside RG being the remnant of the field generated by the convective core dynamo during the main sequence (\textsc{ms}) for M$_\star$ > 1.3 M$_\odot$ (see Fig.~\ref{fig:1}).
We then theoretically investigate the effect of this stable fossil mixed poloidal and toroidal field buried inside the core of RG on their mixed-mode frequencies. In this study, we focus on the axisymmetric case, where the magnetic axis is aligned with the rotation axis of the star. The convective envelope is also represented  by the vortex symbols.\\

\section{Perturbative methods}
\label{sec:3}
We use the linearised equations for stellar oscillations \cite{Unno1989}, perturbed by both rotation and magnetism to evaluate the first-order frequency perturbation $\delta \omega$ through
\begin{equation}
    \delta \omega=-\frac{\langle \fvec{\xi_0},\delta \fvec{F_\ell}(\fvec{\xi_0}) \rangle + \langle \fvec{\xi_0},\fvec{F_c}(\fvec{\xi_0}) \rangle}{2 \omega_0 \langle \fvec{\xi_0}, \fvec{\xi_0} \rangle},
    \label{eq:1}
\end{equation}
with $\delta \fvec{F_\ell}$ being the perturbed volumetric Lorentz force, $\fvec{F_c}$ the Coriolis acceleration, $\fvec{\xi_0}$ the unperturbed eigenfunctions of mixed modes, and $\omega_0$ the unperturbed mixed-mode frequencies. For each of the terms composing Eq.~\ref{eq:1} (see \cite{Prat2019} for more details), we numerically estimate the mass mode two following dominant components

\begin{equation}
\langle \fvec{\xi_0},\fvec{F_c}(\fvec{\xi_0}) \rangle \simeq 8 \pi m \Omega \int_0^R \rho r^2 |\xi_h|^2\textrm{d}r \int_0^\pi Y_\ell^m \partial_\theta Y_\ell^{m*}\frac{\cos{\theta}}{\sin{\theta}}  \textrm{d}\theta,
\label{eq:3}
\end{equation}
\begin{multline}
\langle \fvec{\xi_0},\delta \fvec{F_\ell}(\fvec{\xi_0}) \rangle \simeq 2 \pi B_0^2 \int_0^R r \xi_h^* b_r (r\xi_h b_r)'' \textrm{d}r \\
\times \int_0^\pi\left(\left(\frac{mY_\ell^m}{\sin{\theta}} \right)^2 + (\partial_\theta Y_\ell^m)^2 \right)\cos^2{\theta} \sin{\theta}\textrm{d}\theta,
\label{eq:2}
\end{multline}

and the mode mass
\begin{equation}
\langle \fvec{\xi_0},\fvec{\xi_0} \rangle \simeq 2 \pi \int_0^R \rho r^2 |\xi_h|^2 \textrm{d}r \int_0^\pi\left(\left(\frac{mY_\ell^m}{\sin{\theta}} \right)^2 + (\partial_\theta Y_\ell^m)^2  \right) {\sin{\theta}}  \textrm{d}\theta,
\label{eq:4}
\end{equation}

\noindent
with $B_0$ the magnetic field strength, $\xi_h(r)$ the radial eigenfunction describing the horizontal displacement on the spherical harmonics ($Y_\ell^m(\theta,\varphi)$ where $\ell$ is the degree, m the azimuthal degree, and $\left(r, \theta, \varphi\right)$ are the usual spherical coordinates), $b_r(r)$ the radial function describing the radial component of the magnetic field on the $\ell=1$ dipolar spherical harmonics, $\Omega$ the angular velocity, and $\rho$ the density profile of the star. The unperturbed eigenfunctions of mixed modes ($\fvec{\xi_0}$) are evaluated by using the stellar pulsation code GYRE \citep{Townsend2013} along with the stellar evolution model MESA \citep{Paxton2011}.

\section{Mixed-mode splittings in RG}
\label{sec:3}


\begin{figure}[t]
\includegraphics[scale=0.31]{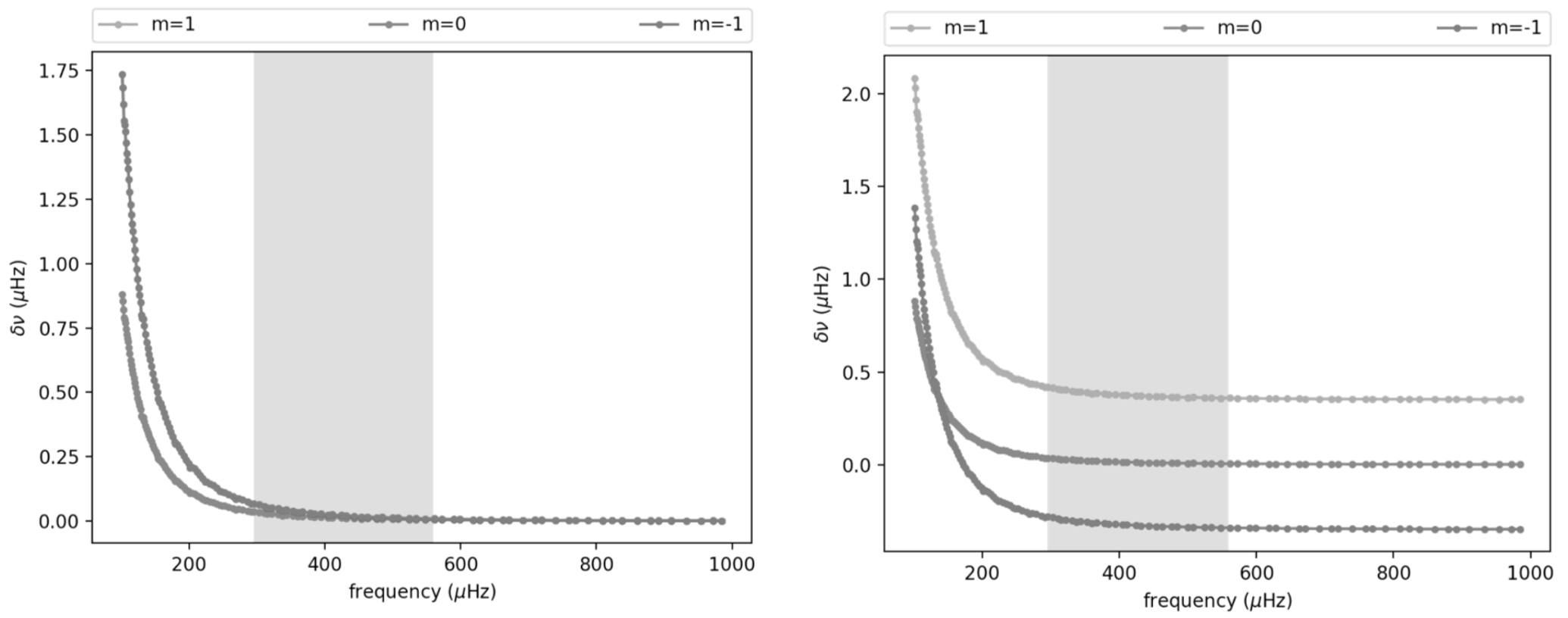}
  \caption{\textsl{Left:} Mode splittings ($\delta {\nu}$) induced by magnetism only for a M$_\star$ = 1.5 M$_\odot$ early RG with $\nu_\textnormal{max}=$ $425\upmu$Hz as a function of mode frequencies. The light gray area represents the range of frequencies $[\nu_\textnormal{max}-5\Delta \nu: \nu_\textnormal{max} +5\Delta \nu]$ for which mixed modes should be visible in real data. Each of them represents a simulated mixed mode with the lower line representing $m=$0 and the upper line $m=$1 and $m=$-1 modes, which overlap as a result of the m dependency in Eq.~\ref{eq:2}. \textsl{Right:} Same as left panel with rotation splitting corresponding to a uniform rotation of 0.7 $\upmu$Hz added. 
  }
  \label{fig:2}       
\end{figure}

Figure~\ref{fig:2} shows the strength of the frequency perturbation of mixed modes due to the magnetic field only (left panel) for a field strength of $B_0=1$ MG. We observe that one individual $\ell=1$ mixed mode splits into two modes: the $m=$0 component and the $m=$1 and $m=$-1 components (overlaping as described by Eq.~\ref{eq:2}). In the frequency region where we expect to see mixed modes in the power spectrum density (light gray area), magnetic splittings are of the order of 0.1 $\upmu$Hz. This is a small effect by comparison with the rotational frequency splitting due to a uniform rotation of the star of $0.7 \upmu$Hz as shown on the right panel. Nevertheless, the effect of such a magnetic field can be large enough to be observable with current instruments.\\

\begin{figure}[t]
\sidecaption[t]
\includegraphics[scale=.72]{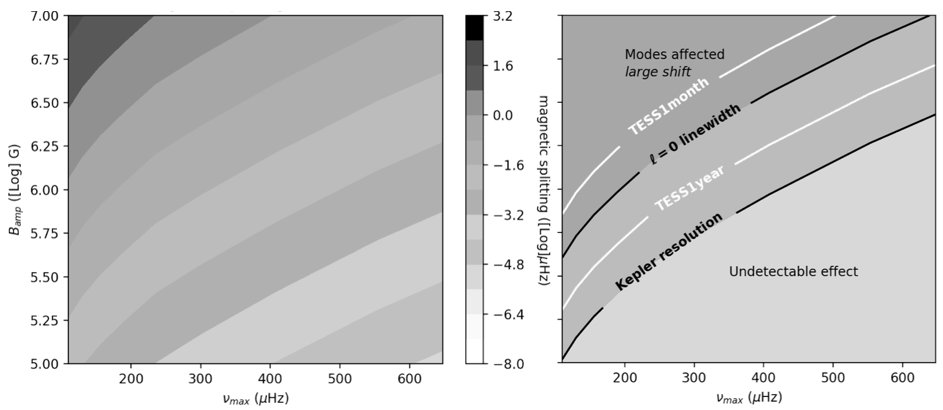}
\caption{\textsl{Left:} Mode splitting induced by magnetism only for a M$_\star$ = 1.5 M$_\odot$ giant as a function of the frequency of maximum power $\nu_\textnormal{max}$ and the magnetic field strength. The mode splitting is calculated at each evolutionary stage by taking the splitting corresponding to the central mixed mode (the closest to $\nu_\textnormal{max}$). \textsl{Right:} Same diagram as the left panel simplified to compare with typical frequencies: the "\textit{Kepler} resolution" line corresponds to the 7.9 nHz resolution for the 4 years \textit{Kepler} data, the "TESS 1 year" line corresponds to the 32 nHz resolution for 1 year of TESS data, the "TESS 1 month" line corresponds to the 380 nHz resolution for 1 month of TESS data, and the "$\ell=0$" line corresponds to the typical radial mode linewidths \citep{Vrard2017} for sub-giants, as being an upper limit for mixed-mode linewidth.}
\label{fig:3}       
\end{figure}

The dependence of the magnetic splittings on the evolution of the star and on the magnetic field strength is represented in Fig.~\ref{fig:3}. The right panel indicates typical data frequency boundaries for the observation of magnetic splittings to be compared to  the left panel on which values of magnetic  splitting are represented: if the combination of magnetic field strength and of the evolution leads to a magnetic splitting below the \textit{Kepler} resolution, then the effect of magnetism on mixed-mode frequencies would not be visible in \textit{Kepler} data. Above the typical $\ell =0$ linewidth the magnetic effect is considered as being too large for our perturbative analysis to be valid (see Bugnet et al., \textsl{in prep}). The region in between the two black lines approximatively shows the ideal conditions for the detection of the magnetic effect in \textit{Kepler} data. This region would be narrowed to the area between the TESS 1 year white line and the $\ell$=0 linewidth black line if considering TESS 1 year data, and we do not expect to be able to detect any magnetic effect on mixed-mode frequencies when looking at only 1 month TESS data.

\section{Conclusion \& Perspectives}

We study the axisymmetric case for which a buried stable fossil magnetic field, with poloidal and toroidal components, is aligned with the rotation axis of the star. We find that a typical field strength of about 1MG results in a detectable frequency shift of the mixed modes inside RG's power spectrum density in addition to the already measured rotational splittings. The study will be extended towards fossil dipolar fields inclined with respect to the rotational axis of the star, and towards non-fossil field topologies such as those derived in \cite{Spruit1999, Jouve2014, Fuller2019}, in order to explore the variety of possible seismic signatures that would allow us to probe the deep magnetism of evolved low-mass stars and its consequences for the transport of angular momentum.\\

\emph{Acknowledgements:} The authors of this work acknowledge the support received from the PLATO CNES grant, the National Aeronautics and Space Administration under Grant NNX15AF13G, by the National Science Foundation grant AST-1411685, the Ramon y Cajal fellowship number RYC-2015-17697, the ERC SPIRE grant (647383), and the Fundation L'Oréal-Unesco-Académie des sciences.


\bibliographystyle{unsrt_7}
 
\end{document}